\documentclass[sigconf]{acmart}
\usepackage{url}
\usepackage{multirow}
\usepackage{balance}
\usepackage{rotating}
\usepackage{booktabs}
\usepackage{hyphsubst}
\usepackage{graphicx}
\usepackage{amsmath}
\graphicspath{{images/}}
\usepackage{hhline}
\usepackage[caption = false]{subfig}

\usepackage{colortbl}

\usepackage{url}
\usepackage{rotating}
\usepackage{amsmath}
\usepackage{cancel}
\usepackage{amssymb}% http://ctan.org/pkg/amssymb
\usepackage{enumitem}
\usepackage{pifont}% http://ctan.org/pkg/pifont

\usepackage{multirow}
\usepackage{array}
\usepackage{placeins}

\usepackage{listings}

\usepackage{color}

\lstdefinestyle{json}
{
    string=[s]{"}{"},
    stringstyle=\color{black},
    comment=[l]{:},
    commentstyle=\color{blue},
}

\makeatletter

\usepackage{algorithm}
\usepackage{algorithmicx}
\usepackage{algpseudocode}
\usepackage{amsmath,amsthm,amssymb}
\usepackage{mathtools}

\DeclarePairedDelimiterX{\infdivx}[2]{(}{)}{%
  #1\;\delimsize|\delimsize|\;#2%
}

\newcommand{\para}[1]{\vspace{2mm}\noindent\textbf{#1}}

\usepackage{rotating}
\usepackage{booktabs}
\usepackage{rotating}
\usepackage{hhline}
\usepackage{hyphsubst}
\usepackage{soul}
\newlength{\Oldarrayrulewidth}

\copyrightyear{2019} 
\acmYear{2019} 
\setcopyright{acmcopyright}
\acmConference{REVEAL@RecSys'2019}{September 16-20, 2019}{Copenhagen, Denmark}
\acmPrice{15.00}
\acmDOI{xx.xxx/xxxxxx.xxxxxx}
\acmISBN{xxx-x-xxxx-xxxx-x/xx/xx}
\begin{document}
\title[Evaluating Tag Recommendations for E-Book Annotation]{Evaluating Tag Recommendations for E-Book Annotation Using a Semantic Similarity Metric}

\author{Emanuel Lacic}
\authornote{Both authors contributed equally to this work.}
\affiliation{%
  \institution{Know-Center GmbH}
  \city{Graz} 
  \state{Austria} 
  }
\email{elacic@know-center.at}

\author{Dominik Kowald}
\authornotemark[1]
\affiliation{%
  \institution{Know-Center GmbH}
  \city{Graz} 
  \state{Austria} 
  }
\email{dkowald@know-center.at}

\author{Dieter Theiler}
\affiliation{%
  \institution{Know-Center GmbH}
  \city{Graz} 
  \state{Austria} 
  }
\email{dtheiler@know-center.at}

\author{Matthias Traub}
\affiliation{%
  \institution{Know-Center GmbH}
  \city{Graz} 
  \state{Austria} 
  }
\email{mtraub@know-center.at}

\author{Lucky Kuffer}
\affiliation{%
  \institution{HGV GmbH}
  \city{Munich} 
  \state{Germany} 
  }
\email{lucky.kuffer@hgv-online.de}

\author{Stefanie Lindstaedt}
\affiliation{%
  \institution{Know-Center GmbH}
  \city{Graz} 
  \state{Austria} 
}
\email{slind@know-center.at}

\author{Elisabeth Lex}
\affiliation{%
  \institution{Graz University of Technology}
  \city{Graz} 
  \state{Austria} 
  }
\email{elisabeth.lex@tugraz.at}

\renewcommand{\shortauthors}{Lacic, E. \& Kowald, D., et al.}

\begin{abstract}
In this paper, we present our work to support publishers and editors in finding descriptive tags for e-books through tag recommendations. We propose a hybrid tag recommendation system for e-books, which leverages search query terms from Amazon users and e-book metadata, which is assigned by publishers and editors. Our idea is to mimic the vocabulary of users in Amazon, who search for and review e-books, and to combine these search terms with editor tags in a hybrid tag recommendation approach. In total, we evaluate 19 tag recommendation algorithms on the review content of Amazon users, which reflects the readers' vocabulary. Our results show that we can improve the performance of tag recommender systems for e-books both concerning tag recommendation accuracy, diversity as well as a novel semantic similarity metric, which we also propose in this paper.
\end{abstract}


%\ccsdesc[500]{Information systems~Social recommendation}

\maketitle

\section{Introduction}
\label{sec:introduction}
When people shop for books online in e-book stores such as, e.g., the Amazon Kindle store, they enter search terms with the goal to find e-books that meet their preferences. Such e-books have a variety of metadata such as, e.g., title, author or keywords, which can be used to retrieve e-books that are relevant to the query. As a consequence, from the perspective of e-book publishers and editors, annotating e-books with tags that best describe the content and which meet the vocabulary of users (e.g., when searching and reviewing e-books) is an essential task~\cite{zubiaga2011tags}.

\para{Problem and aim of this work.} Annotating e-books with suitable tags is, however, a complex task as users' vocabulary may differ from the one of editors. Such a vocabulary mismatch yet hinders effective organization and retrieval~\cite{zhao2010termnecessity} of e-books. For example, while editors mostly annotate e-books with descriptive tags that reflect the book's content, Amazon users often search for parts of the book title. In the data we use for the present study (see Section~\ref{s:method}), we find that around 30\% of the Amazon search terms contain parts of e-book titles. 

In this paper, we present our work to support editors in the e-book annotation process with tag recommendations~\cite{jaschke2007tag,kowald2017tagrec}. Our idea is to exploit user-generated search query terms in Amazon to mimic the vocabulary of users in Amazon, who search for e-books. We combine these search terms with tags assigned by editors in a hybrid tag recommendation approach. 
Thus, our aim is to show that we can improve the performance of tag recommender systems for e-books both concerning recommendation accuracy as well as semantic similarity and tag recommendation diversity.

\para{Related work.}
In tag recommender systems, mostly content-based algorithms (e.g.,~\cite{lu2009content,lops2013content}) are used to recommend tags to annotate resources such as e-books. 
%\hl{\textbf{Do we really "extend" the type of algorithms?:}} We extend this type of algorithms and 
In our work, we incorporate both content features of e-books (i.e., title and description text) as well as Amazon search terms to account for the vocabulary of e-book readers. 
%\hl{\textbf{Previous sentence reads as: TF-IDF similarities --> vocabulary of e-book readers}}

Concerning the evaluation of tag recommendation systems, 
most studies focus on measuring the accuracy of tag recommendations (e.g.,~\cite{jaschke2007tag}). However, the authors of~\cite{belem2013topic} suggest also to use beyond-accuracy metrics such as diversity to evaluate the quality of tag recommendations. In our work, we measure recommendation diversity in addition to recommendation accuracy and propose a novel metric termed semantic similarity to validate semantic matches of tag recommendations. 

\para{Approach and findings.} We exploit editor tags and user-generated search terms as input for tag recommendation approaches. Our evaluation comprises of a rich set of 19 different algorithms to recommend tags for e-books, which we group into (i) popularity-based, (ii) similarity-based (i.e., using content information), and (iii) hybrid approaches.
We evaluate our approaches in terms of accuracy, semantic similarity and diversity on the review content of Amazon users, which reflects the readers' vocabulary. With semantic similarity, we measure how semantically similar (based on learned Doc2Vec~\cite{le2014distributed} embeddings) the list of recommended tags is to the list of relevant tags. We use this additional metric to measure not only exact ``hits'' of our recommendations but also semantic matches.

Our evaluation results show that combining both data sources enhances the quality of tag recommendations for annotating e-books. Furthermore, approaches that solely train on Amazon search terms provide poor performance in terms of accuracy but deliver good results in terms of semantic similarity and recommendation diversity.

%%%%%%%%%%%%%%%%%%%%%%%%%%%%%%%%%%%%%%%%%%
\section{Method} \label{s:method}
In this section, we describe our dataset as well as our tag recommendation approaches we propose to annotate e-books. %\hl{\textbf{should be our approaches instead of our dataset}}

\subsection{Dataset} \label{s:dataset}
Our dataset contains two sources of data, one to generate tag recommendations and another one to evaluate tag recommendations. HGV GmbH has collected all data sources\footnote{Currently the data used in this study cannot be made publicly available because of copyright issues, but we will try to provide a public version of it soon in the future.} and we provide the dataset statistics in Table~\ref{tab:datasets}.

\para{Data used to generate recommendations.} 
We employ two sources of e-book annotation data: (i) editor tags, and (ii) Amazon search terms. For editor tags, we collect data of 48,705 e-books from 13 publishers, namely Kunstmann, Delius-Klasnig, VUR, HJR, Diogenes, Campus, Kiwi, Beltz, Chbeck, Rowohlt, Droemer, Fischer and Neopubli. Apart from the editor tags, this data contains metadata fields of e-books such as the ISBN, the title, a description text, the author and a list of BISACs, which are identifiers for book categories.

For the Amazon search terms, we collect search query logs of 21,243 e-books for 12 months (i.e., November 2017 to October 2018). Apart from the search terms, this data contains the e-books' ISBNs, titles and description texts.

Table~\ref{tab:datasets} shows that the overlap of e-books that have editor tags and Amazon search terms is small (i.e., only 497). Furthermore, author and BISAC (i.e., the book category identifier) information are primarily available for e-books that contain editor tags. Consequently, both data sources provide complementary information, which underpins the intention of this work, i.e., to evaluate tag recommendation approaches using annotation sources from different contexts.

\para{Data used to evaluate recommendations.}
For evaluation, we use a third set of e-book annotations, namely Amazon review keywords. These review keywords are extracted from the Amazon review texts and are typically provided in the review section of books on Amazon. Our idea is to not favor one or the other data source (i.e., editor tags and Amazon search terms) when evaluating our approaches against expected tags. At the same time, we consider Amazon review keywords to be a good mixture of editor tags and search terms as they describe both the content and the users' opinions on the e-books (i.e., the readers' vocabulary). As shown in Table~\ref{tab:datasets}, we collect Amazon review keywords for 2,896 e-books (publishers: Kiwi, Rowohlt, Fischer, and Droemer), which leads to 33,663 distinct review keywords and on average 30 keyword assignments per e-book.

\begin{table}[t!]
    \small
  \setlength{\tabcolsep}{4.0pt}    
  \centering    
    \begin{tabular}{l|r}
    \specialrule{.2em}{.1em}{.1em}
\textbf{Data used to generate recommendations} & \textbf{\#} \\\hline
                                            Number of e-books                                        & 69,451            \\
                                            thereof with editor tags                                        &    48,705            \\
                                            thereof with Amazon search terms                        & 21,243              \\
                                            thereof with editor tags and Amazon search terms    & 497              \\
                                            Number of distinct authors                                & 25,086                    \\
                                            Number of distinct BISACs (= category IDs)         & 1,448                \\
                                            Number of distinct editor tags                    & 114,707                     \\
                                            Number of distinct Amazon search terms                    & 8,240                  \\
        \specialrule{.2em}{.1em}{.1em}    
\textbf{Data used to evaluate recommendations} & \textbf{\#} \\\hline
Number of e-books with Amazon review keywords        & 2,896            \\
Number of distinct Amazon review keywords & 33,663 \\
Avg. number of distinct Amazon review keywords per e-book & 30 \\
        \specialrule{.2em}{.1em}{.1em}                                
    \end{tabular}
      \caption{Statistics of our e-book annotation dataset used to generate and evaluate tag recommendations.
      \vspace{-7mm}}    
  \label{tab:datasets}
\end{table}

\subsection{Tag Recommendation Approaches}
\label{sec:approaches}
We implement three types of tag recommendation approaches, i.e., (i) popularity-based, (ii) similarity-based (i.e., using content information), and (iii) hybrid approaches. Due to the lack of personalized tags (i.e., we do not know which user has assigned a tag), we do not implement other types of algorithms such as collaborative filtering~\cite{marinho2008collaborative}. In total, we evaluate 19 different algorithms to recommend tags for annotating e-books.

\vspace{2mm}
\noindent 
\textbf{Popularity-based approaches.} 
We recommend the most frequently used tags in the dataset, which is a common strategy for tag recommendations~\cite{kowald2016influence}. That is, a most popular $MP_{Editor}$ approach for editor tags and a most popular $MP_{Amazon}$ approach for Amazon search terms. For e-books, for which we also have author (= $MP^{Author}_{Editor}$ and $MP^{Author}_{Amazon}$) or BISAC (=$MP^{BISAC}_{Editor}$ and $MP^{BISAC}_{Amazon}$) information, we use these features to further filter the recommended tags, i.e., to only recommend tags that were used to annotate e-books of a specific author or a specific BISAC.
%\elex{not completely clear what is filtered and with what purpose}

We combine both data sources (i.e., editor tags and Amazon search terms) using a round-robin combination strategy, which ensures an equal weight for both sources. This gives us three additional popularity-based algorithms (= $MP_{Combined}$, $MP^{Author}_{Combined}$ and $MP^{BISAC}_{Combined}$).

\vspace{2mm}
\noindent 
\textbf{Similarity-based approaches.} We exploit the textual content of e-books (i.e., description or title) 
%\elex{mention already earlier that we also use text content as input} 
to recommend relevant tags~\cite{cantador2010content}. For this, we first employ a content-based filtering approach~\cite{balabanovic1997fab} based on TF-IDF~\cite{ramos2003using} to find top-$N$ similar e-books\footnote{In our experiments, we set $N = 20$, the minimum document frequency to $10$ and the minimum word length to $5$.}. For each of the similar e-books, we then either extract the assigned editor tags (= $SIM^{Description}_{Editor}$ and $SIM^{Title}_{Editor}$) or the Amazon search terms (= $SIM^{Description}_{Amazon}$ and $SIM^{Title}_{Amazon}$). To combine the tags of the top-$N$ similar e-books, we use the cross-source algorithm~\cite{bostandjiev2012tasteweights}, which favors tags that were used to annotate more than one similar e-book (i.e., tags that come from multiple recommendation sources). 
The final tag relevancy is calculated as:

  \begin{align}
        W_{t_i} =  |S_{t_i}| \cdot \sum\limits_{s_{t_i} \in S}{W_{s_{t_i}}}
  \end{align}

where $|S_{t_i}|$ denotes the number of distinct e-books, which yielded the recommendation of tag $t_{i}$, to favor tags that come from multiple sources and $W_{s_{t_i}}$ is the similarity score of the corresponding e-book. 
We again use a round-robin strategy to combine both data sources (= $SIM^{Description}_{Combined}$ and $SIM^{Title}_{Combined}$).

\vspace{2mm}
\noindent 
\textbf{Hybrid approaches.} We use the previously mentioned cross-source algorithm~\cite{bostandjiev2012tasteweights} to construct four hybrid recommendation approaches. 
In this case, tags are favored that are recommended by more than one algorithm.

Hence, to create a popularity-based hybrid (= $HYB^{MP}$), we combine the best three performing popularity-based approaches from the ones (i) without any contextual signal, (ii) with the author as context, and (iii) with BISAC as context. In the case of the similarity-based hybrid (= $HYB^{SIM}$), we utilize the two best performing similarity-based approaches from the ones (i) which use the title, and (ii) which use the description text. We further define $HYB^{All}$, a hybrid approach that combines the three popularity-based methods of $HYB^{MP}$ and the two similarity-based approaches of $HYB^{SIM}$. Finally, we define $HYB^{Best}$ as a hybrid approach that uses the best performing popularity-based and the best performing similarity-based approach (see Figure~\ref{fig:accuracy_res} in Section~\ref{sec:res} for more details about the particular algorithm combinations).

\section{Experimental Setup} \label{sec:exp}
In this section, we describe our evaluation protocol as well as the measures we use to evaluate and compare our tag recommendation approaches.

\subsection{Evaluation Protocol}
For evaluation, we use the third set of e-book annotations, namely Amazon review keywords. As described in Section~\ref{s:dataset}, these review keywords are extracted from the Amazon review texts and thus, reflect the users' vocabulary. We evaluate our approaches for the 2,896 e-books, for whom we got review keywords. To follow common practice for tag recommendation evaluation~\cite{kowald2015evaluating}, we predict the assigned review keywords (= our test set) for respective e-books.

\subsection{Evaluation Metrics}
\label{sec:method}

In this work, we measure (i) recommendation accuracy, (ii) semantic similarity, and (iii) recommendation diversity to evaluate the quality of our approaches from different perspectives.

\vspace{2mm}
\noindent 
\textbf{Recommendation accuracy.} We use \textit{Normalized Discounted Cumulative Gain} (nDCG)~\cite{ParraSahebi} to measure the accuracy of the tag recommendation approaches. The nDCG measure is a standard ranking-dependent metric that not only measures how many tags can be correctly predicted but also takes into account their position in the recommendation list with length of $k$. 
It is based on the \textit{Discounted Cummulative Gain}, which is given by: 

   \begin{align}
      DCG@k = \sum\limits_{k = 1}\limits^{|r_b^k|} (\frac{2 ^ {T(k)} - 1}{log_{2}(1 + k)})
   \end{align}
   
where $T(k)$ is a function that returns 1 if the recommended tag at position $i$ in the recommended list is relevant. We then calculate DCG@$k$ for every evaluated e-book by dividing DCG@$k$ with the ideal DCG value iDCG@$k$, which is the highest possible DCG value that can be achieved if all the relevant tags would be recommended in the correct order. It is given by the following formula~\cite{ParraSahebi}:
  
  \begin{align}
      nDCG@k = \frac{ 1 }{ |B| } \sum\limits_{ b \in B }{ (\frac{DCG@k}{iDCG@k}) }
\end{align}

\vspace{2mm}
\noindent 
\textbf{Semantic similarity.}
One precondition of standard recommendation accuracy measures is that to generate a ``hit'', the recommended tag needs to be an exact syntactical match to the one from the test set. When tags are recommended from one data source and compared to tags from another source, this can be problematic. 
For example, if we recommend the tag ``victim'' but expect the tag ``prey'', we would mark this as a mismatch, therefore being a bad recommendation. But if we know that the corresponding e-book is a crime novel, the recommended tag would be (semantically) descriptive to reflect the book's content. Hence, in this paper, we propose to additionally measure the semantic similarity between recommended tags and tags from the test set (i.e., the Amazon review keywords).

\begin{figure*}[t!]
\centering
\subfloat[Popularity-based approaches]{
  \centering
  \includegraphics[width=.33\textwidth]{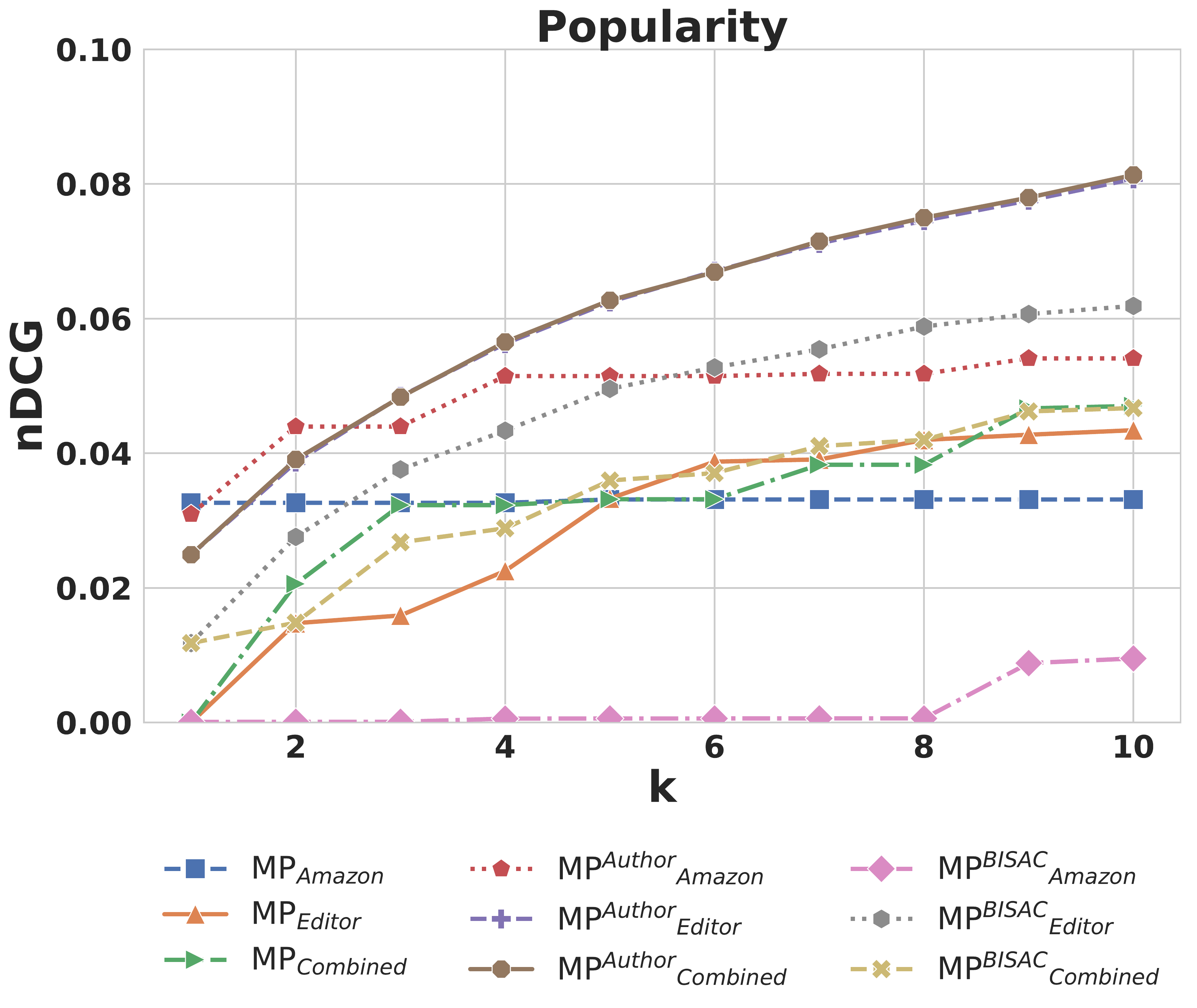}
  \label{fig:accuracy_res_pop}
} 
\subfloat[Similarity-based approaches]{
  \centering
  \includegraphics[width=0.33\textwidth]{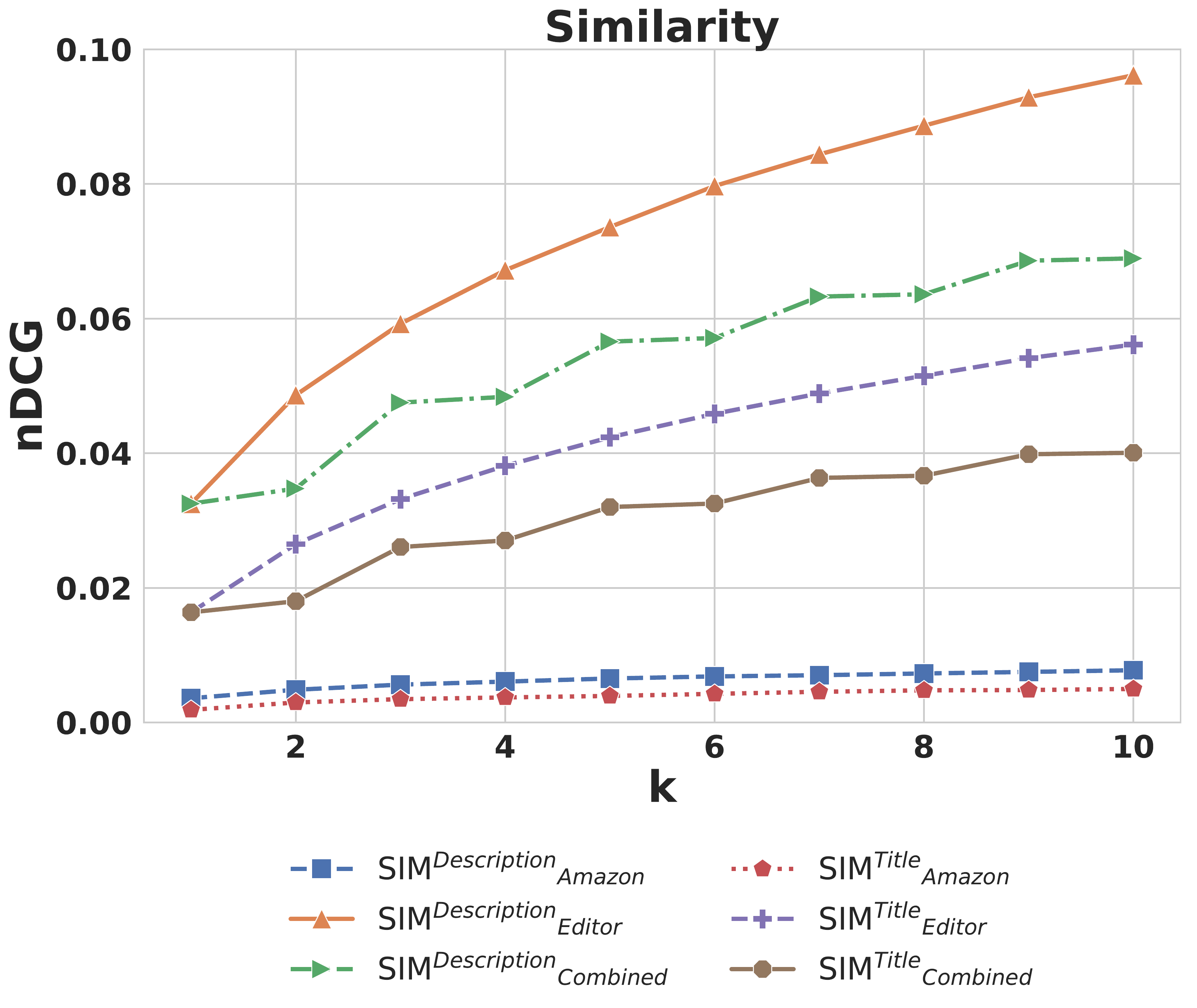}
    \label{fig:accuracy_res_sim}
} 
\subfloat[Hybrid approaches]{
  \centering
  \includegraphics[width=.33\textwidth]{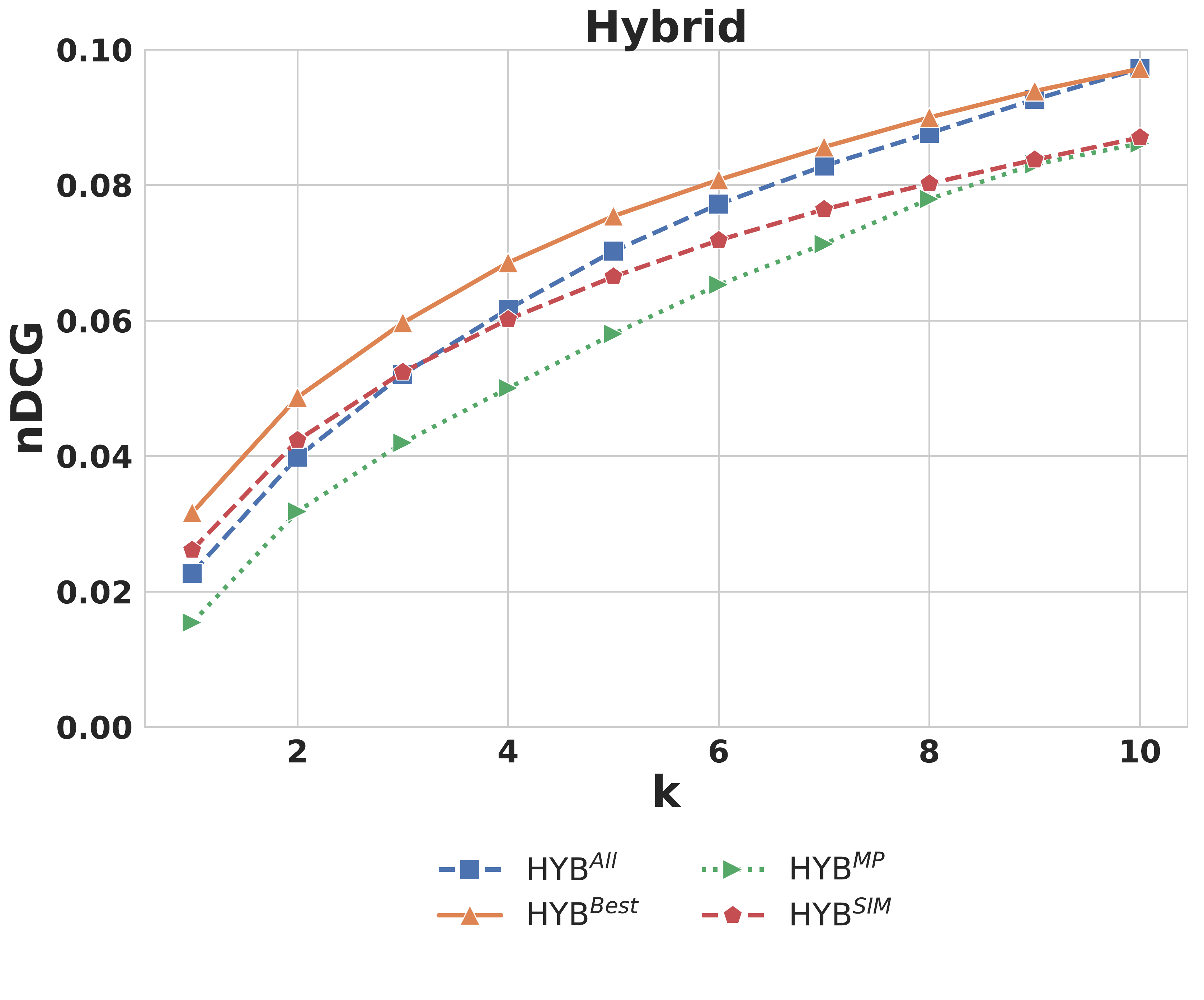}
    \label{fig:accuracy_res_hyb}
} 
\caption{Accuracy results with respect to $nDCG$ for (a) popularity-based, (b) similarity-based and (c) hybrid tag recommendation approaches. All results are reported for different numbers of recommended tags (i.e., $k \in [1,10]$).\vspace{-3mm}}
\label{fig:accuracy_res}
\end{figure*}

Over the last four years, there have been several notable publications in the area of applying deep learning to uncover semantic relationships between textual content (e.g., by learning word embeddings with Word2Vec~\cite{mikolov2013distributed,kenter2015short}).  
Based on this, we propose an alternative measure of recommendation quality by learning the semantic relationships from both vocabularies and then using it to compare how semantically similar the recommended tags are to the expected review keywords. %\hl{\textbf{maybe rework last sent}}
For this, we first extract the textual content in the form of the description text, title, editor tags and Amazon search terms of e-books from our dataset\footnote{We also pre-process the extracted text by removing bad characters, stop words and changing all words to lowercase.}. We then train a Doc2Vec \cite{le2014distributed} model\footnote{We use the DBOW approach with a size of $50$ for the latent vector representation, negative sampling of $10$, a learning rate of $0.025$ and train it for $10$ epochs.} on the content.
Then, we use the model to infer the latent representation for both the complete list of recommended tags as well as the list of expected tags from the test set. Finally, we use the cosine similarity measure to calculate how semantically similar these two lists are. %\hl{\textbf{ means not on per tag basis?}}

\vspace{2mm}
\noindent 
\textbf{Recommendation diversity.}
As defined in~\cite{SmythMcClave01}, we calculate recommendation diversity as the average dissimilarity of all pairs of tags in the list of recommended tags. Thus, given a distance function $d(t_i,t_j)$ that corresponds to the dissimilarity between two tags $t_i$ and $t_j$ in the list of recommended tags, $D$ is given as the average dissimilarity of all pairs of tags: 

  \begin{align}
        D@k = \frac{ 1 }{ |B| } \sum\limits_{ b \in B }{(\frac{ 1 }{ k \cdot (k - 1) } \sum\limits_{ i \in R }{ \sum\limits_{ j \in r_b^k, j \neq i  }{  }{ d(t_i,t_j) } })}
    \end{align}
where $|B|$ is the number of evaluated e-books and the dissimilarity function is defined as $d(t_i,t_j) = 1 - sim(t_i,t_j)$. 
In our experiments, we use the previously trained Doc2Vec model to extract the latent representation of a specific tag.
The similarity of two tags $sim(t_i,t_j)$ is then calculated with the Cosine similarity measure using the latent vector representations of respective tags $t_i$ and $t_j$.

\section{Results} \label{sec:res}
Concerning tag recommendation accuracy, in this section, we report results for different values of $k$ (i.e., number of recommended tags). For the beyond-accuracy experiment, we use the full list of recommended tags (i.e., $k=10$).

\subsection{Recommendation Accuracy Evaluation}
Figure~\ref{fig:accuracy_res} shows the results of the accuracy experiment for the (i) popularity-based, (ii) similarity-based, and (iii) hybrid tag recommendation approaches.

\vspace{2mm}
\noindent 
\textbf{Popularity-based approaches.} In Figure~\ref{fig:accuracy_res_pop}, we see that popularity-based approaches based on editor tags tend to perform better than if trained on Amazon search terms. If we take into account contextual information like BISAC or author, we can further improve accuracy in terms of $nDCG$. 
That is, we find that using popular tags from e-books of a specific author leads to the best accuracy of the popularity-based approaches. This suggests that editors and readers do seem to reuse tags for e-books of same authors.
%(i.e., $nDCG@10 = 0.0807$ for $MP^{Author}_{Editor}$). 
%\hl{\textbf{not seen in figure, should different one?}} 
If we use both editor tags and Amazon search terms, we can further increase accuracy, especially for higher values of $k$ like in the case of $MP_{Combined}$. This is, however, not the case for $MP^{BISAC}_{Combined}$ as the accuracy of the integrated $MP^{BISAC}_{Amazon}$ approach is low. The reason for this is the limited amount of e-books from within the Amazon search query logs that have BISAC information (i.e., only $2.38\%$).

\vspace{2mm}
\noindent 
\textbf{Similarity-based approaches.} We further improve accuracy if we first find similar e-books and then extract their top-$k$ tags in a cross-source manner as described in Section~\ref{sec:approaches}.

As shown in Figure~\ref{fig:accuracy_res_sim}, using the description text to find similar e-books results in more accurate tag recommendations than using the title (i.e., $nDCG@10 = 0.0961$ for $SIM^{Description}_{Editor}$). This is somehow expected as the description text consists of a bigger corpus of words (i.e., multiple sentences) than the title. Concerning the collected Amazon search query logs, extracting and then recommending tags from this source results in a much lower accuracy performance. Thus, these results also suggest to investigate beyond-accuracy metrics as done in Section~\ref{s:beyond}.
%\hl{\textbf{more explanation would be interesting}}

\begin{figure*}[t!]
\centering
\subfloat[Semantic similarity]{
  \centering
  \includegraphics[width=.97\columnwidth]{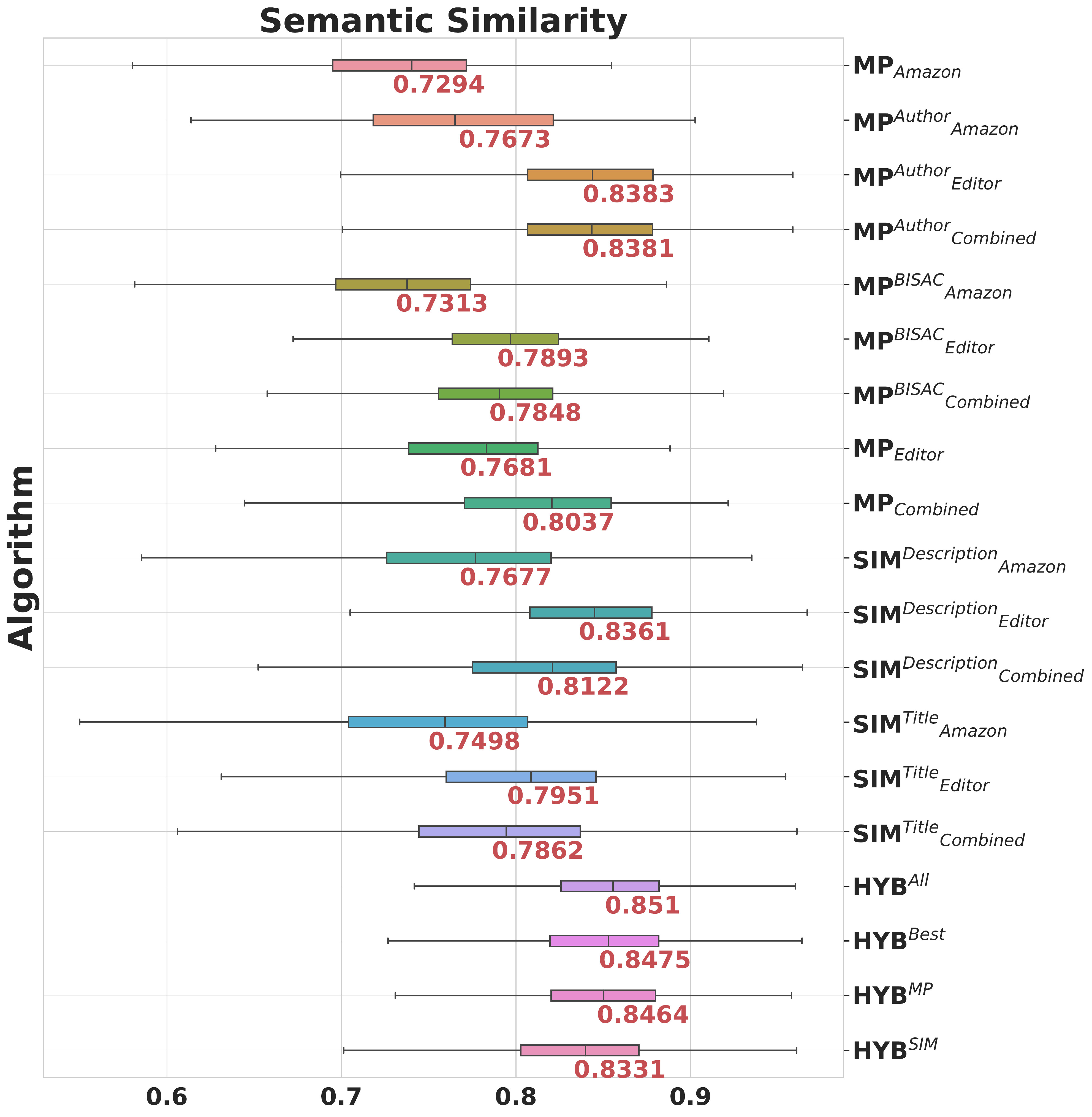}
    \label{fig:similarity}
} 
\subfloat[Recommendation diversity]{
  \centering
  \includegraphics[width=.97\columnwidth]{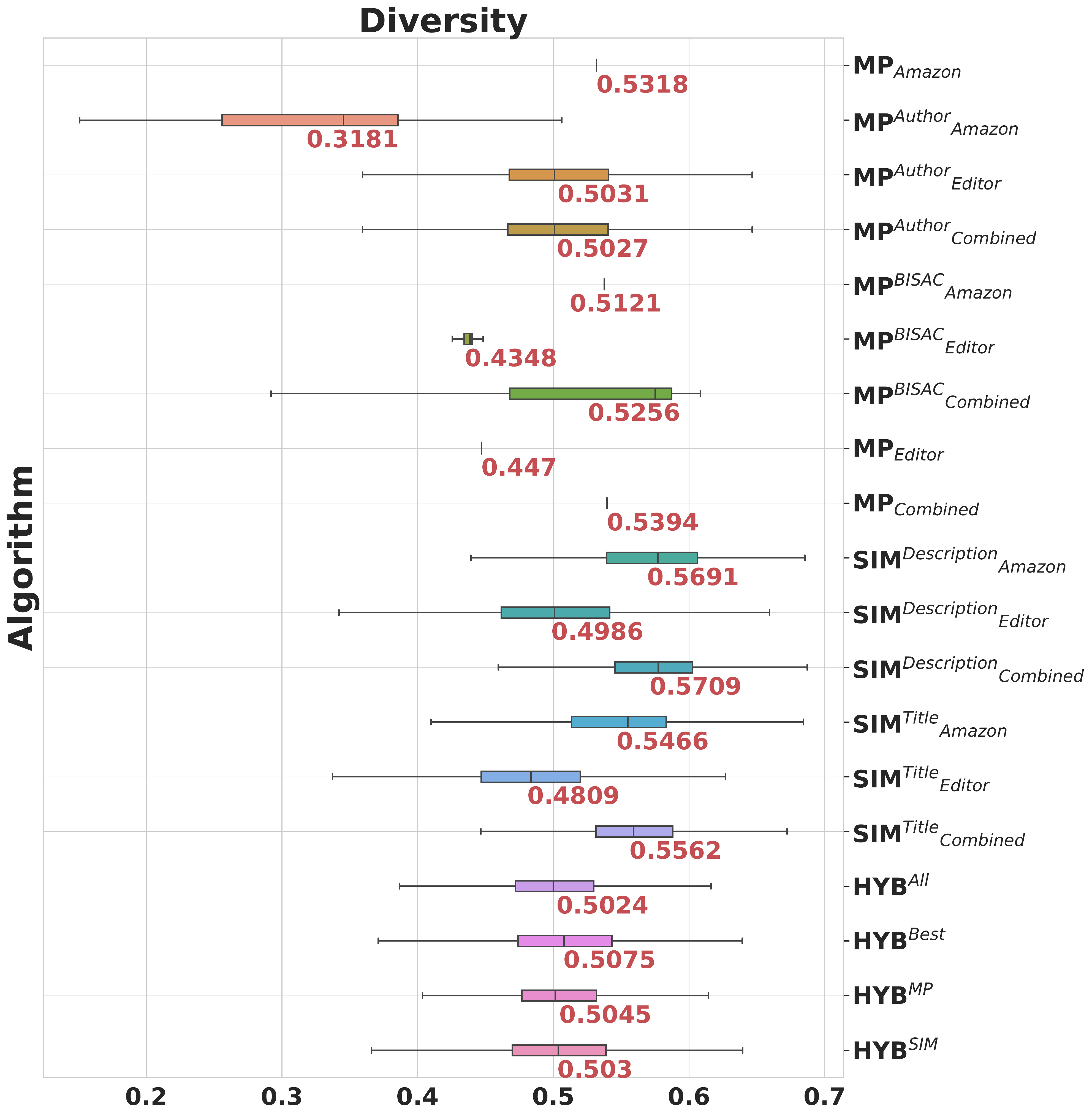}
  \label{fig:diversity}
} 
\caption{Beyond-accuracy evaluation results of our tag recommendation approaches. We use the list of $10$ recommended tags to calculate the (a) semantic similarity, and (b) recommendation diversity. We provide the boxplots and the mean values for the approaches.\vspace{-3mm}}
\label{fig:div_sim}
\end{figure*}

\vspace{2mm}
\noindent 
\textbf{Hybrid approaches.} Figure~\ref{fig:accuracy_res_hyb} shows the accuracy results of the four hybrid approaches. By combining the best three popularity-based approaches, we outperform all of the initially evaluated popularity algorithms (i.e., $nDCG@10 = 0.0862$ for $HYB^{MP}$). On the contrary, the combination of the two best performing similarity-based approaches $SIM^{Description}_{Editor}$ and $SIM^{Title}_{Editor}$ does not yield better accuracy. The negative impact of using a lower-performing approach such as $SIM^{Title}_{Editor}$ within a hybrid combination can also be observed in $HYB^{All}$ for lower values of $k$. Overall, this confirms our initial intuition that combining the best performing popularity-based approach with the best similarity-based approach should result in the highest accuracy (i.e., $nDCG@10 = 0.0972$ for $HYB^{Best}$). Moreover, our goal, namely to exploit editor tags in combination with search terms used by readers to increase the metadata quality of e-books, is shown to be best supported by applying hybrid approaches as they provide the best prediction results.

\subsection{Beyond-Accuracy Evaluation} \label{s:beyond}
Figure~\ref{fig:div_sim} illustrates the results of the experiments, which measure the recommendation impact beyond-accuracy.

\vspace{2mm}
\noindent 
\textbf{Semantic similarity.} Figure~\ref{fig:similarity} illustrates the results of our proposed semantic similarity measure. To compare our proposed measure to standard accuracy measures such as $nDCG$, we use Kendall's Tau rank correlation \cite{kendall1938new} as suggested by \cite{lapata2006automatic} for automatic evaluation of information-ordering tasks. From that, we rank our recommendation approaches according to both accuracy and semantic similarity and calculate the relation between both rankings. This results in $\tau = 0.743$ with a p-value < $0.00001$, which suggests a high correlation between the semantic similarity and the standard accuracy measure. 

Therefore, the semantic similarity measure helps us interpret the recommendation quality. 
%\hl{\textbf{why besides? first paragraph was only an elaboration of the feasibility of comparison, right?}}
For instance, we achieve the lowest $nDCG$ values with the similarity-based approaches that recommend Amazon search terms (i.e., $SIM^{Description}_{Amazon}$ and $SIM^{Title}_{Amazon}$). %\hl{\textbf{guess, sentence doesn't mean what it says}}
When comparing these results with others from Figure~\ref{fig:accuracy_res_sim}, a conclusion could be quickly drawn that the recommended tags are merely unusable. However, by looking at Figure~\ref{fig:similarity}, we see that, although these approaches do not provide the highest recommendation accuracy, they still result in tag recommendations that are semantically related at a high degree\footnote{A semantic similarity of $1.0$ would denote a semantically (and syntactically) perfect fit of tag recommendations to the test set.} to the expected annotations from the test set. Overall, this suggests that approaches, which provide a poor accuracy performance concerning $nDCG$ but provide a good performance regarding semantic similarity could still be helpful for annotating e-books. 

\vspace{2mm}
\noindent 
\textbf{Recommendation diversity.} Figure~\ref{fig:diversity} shows the diversity of the tag recommendation approaches. We achieve the highest diversity with the similarity-based approaches, which extract Amazon search terms. Their accuracy is, however, very low. Thus, the combination of the two vocabularies can provide a good trade-off between recommendation accuracy and diversity.

\section{Conclusion and Future Work} \label{sec:conc}
In this paper, we present our work to support editors in the e-book annotation process. Specifically, we aim to provide tag recommendations that incorporate both the vocabulary of the editors and e-book readers. Therefore, we train various configurations of tag recommender approaches on editors' tags and Amazon search terms and evaluate them on a dataset containing Amazon review keywords. We find that combining both data sources enhances the quality of tag recommendations for annotating e-books. Furthermore, while approaches that train only on Amazon search terms provide poor performance concerning recommendation accuracy, we show that they still offer helpful annotations concerning recommendation diversity as well as our novel semantic similarity metric.

\vspace{2mm}
\noindent 
\textbf{Future work.}
For future work, we plan to validate our findings using another dataset, e.g., by recommending tags for scientific articles and books in BibSonomy. With this, we aim to demonstrate the usefulness of the proposed approach in a similar domain and to enhance the reproducibility of our results by using an open dataset. 

Moreover, we plan to evaluate our tag recommendation approaches in a study with domain users. Also, we want to improve our similarity-based approaches by integrating novel embedding approaches~\cite{mikolov2013distributed,kenter2015short} as we did, for example, with our proposed semantic similarity evaluation metric. Finally, we aim to incorporate explanations for recommended tags so that editors of e-book annotations receive additional support in annotating e-books~\cite{vig2009tagsplanations}.
By making the underlying (semantic) reasoning visible to the editor who is in charge of tailoring annotations, we aim to support two goals: (i) allowing readers to discover e-books more efficiently, and (ii) enabling publishers to leverage semi-automatic categorization processes for e-books. In turn, providing explanations fosters control over which vocabulary to choose when tagging e-books for different application contexts.

\vspace{2mm}
\noindent 
\textbf{Acknowledgments.} The authors would like to thank Peter Langs, Jan-Philipp Wolf and Alyona Schraa from HGV GmbH for providing the e-book annotation data. This work was funded by the Know-Center GmbH (FFG COMET Program), the FFG Data Market Austria project and the AI4EU project (EU grant 825619). The Know-Center GmbH is funded within the Austrian COMET Program - Competence Centers for Excellent Technologies - under the auspices of the Austrian Ministry of Transport, Innovation and Technology, the Austrian Ministry of Economics and Labor and by the State of Styria. COMET is managed by the Austrian Research Promotion Agency (FFG).

%\newpage{}
\balance

\bibliographystyle{abbrv}
%\small

%\bibliography{umap19}

\end{document}